\documentclass{article}
\usepackage{colm2024_conference}

\usepackage{microtype}
\usepackage{hyperref}
\usepackage{url}
\usepackage{booktabs}   
\usepackage{soul}
\usepackage[T1]{fontenc}
\usepackage{listings}

\definecolor{darkblue}{rgb}{0, 0, 0.5}
\definecolor{typhoonpurple}{RGB}{113, 107, 216}
\hypersetup{colorlinks=true, citecolor=darkblue, linkcolor=darkblue, urlcolor=darkblue}

\usepackage{graphicx}
\usepackage{amsmath,amssymb,amsfonts}
\usepackage{algorithmic}
\usepackage{textcomp}
\usepackage{xcolor}

\usepackage{color, tabularx}
\usepackage{booktabs, multirow}
\usepackage{float}
\usepackage{url}
\usepackage{color, colortbl}
\definecolor{Gray}{gray}{0.9}
\usepackage{float}
\usepackage{caption}
\usepackage{diagbox}
\usepackage{hyperref}

\usepackage{pifont}

\usepackage{cleveref}

\NewDocumentCommand\emojilogo{}{
\includegraphics[scale=0.35,trim=0cm 2cm 0 0]{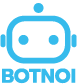}
}

\title{%
\begin{tabular}{@{}m{0.08\textwidth}@{\hspace{0.02\textwidth}}p{0.88\textwidth}@{}}
\emojilogo & Domain-Adaptive ASR for Telephony AI Agents: Fine-tuning Canary Flash Models for Enterprise Contact Center Applications
\end{tabular}}

\author{\vspace{-4mm} \\
Chanameth Boonpramuk \\
Botnoi Group \\
\texttt{chanameth.b@botnoigroup.com} \\
\And
{\mdseries Winn Voravuthikunchai} \\
Botnoi Group \\
\texttt{winnv@botnoigroup.com} \\
\And
{\mdseries Songpol Bunyang} \\
Botnoi Group \\
\texttt{songpol.b@botnoigroup.com} \\
}

\colmfinalcopy
\begin{document}

\maketitle

\begin{abstract}
This technical report describes Botnoi Group's methodology and results for rapidly fine-tuning the open-source NVIDIA Canary 180M Flash and NVIDIA Canary 1B Flash multitask models for speech-to-text tasks using the NVIDIA NeMo framework, with a focus on telephony-grade audio. To support this adaptation, we construct a telephony-oriented fine-tuning dataset from live voicebot system recordings and prompted speech with telephony-oriented augmentation. We evaluate four targeted experiments---language adaptation (Thai), telephony robustness, domain-specific jargon (names and addresses), and latency---using character error rate (CER) for accuracy and real-time factor (RTFx) for inference speed. Results show that fine-tuning substantially improves recognition in noisy telephony environments, reducing CER from 23.31\% to 9.04\% on BOTNOI telephony data, and further improves business-critical names and addresses from 16.98\% to 3.78\% CER through domain-specific adaptation. Overall, our results show that domain-adaptive fine-tuning enhances business-critical terminology while preserving real-time responsiveness for production voicebot deployments, and the BOTNOI ASR telephony benchmark is publicly available at \url{https://github.com/OHM-Songpol/Botnoi_ASR_telephony}.
\end{abstract}

\section{Introduction}

Automatic Speech Recognition (ASR) has become a foundational technology for real-time voice-based AI systems, powering applications that range from virtual assistants and meeting transcription to enterprise contact center automation~\citep{hinton2012deep,amodei2016deepspeech2,whisper2022}. At Botnoi Group, we develop AI agent solutions---most notably voicebot systems---that support customer interactions across Thailand and Southeast Asia. A key requirement for these systems is an ASR engine that achieves both high accuracy and low latency under real-world operating conditions, where background noise, channel distortion, and speaker variability are the norm rather than the exception~\citep{li2014robust}.

Despite significant progress driven by end-to-end neural architectures~\citep{graves2006connectionist,graves2012rnnt,chan2016las} and large-scale pretraining~\citep{baevski2020wav2vec2,whisper2022}, off-the-shelf ASR models often underperform in telephony environments due to a mismatch between training and deployment conditions. Most ASR systems are trained predominantly on 16~kHz wideband audio, whereas telephone speech is typically 8~kHz narrowband and subject to codec compression, packet loss, and handset noise, leading to degraded performance in real-world call scenarios~\citep{li2012mixedbandwidth,shahnawazuddin2019codec}. These challenges are further amplified in low-resource languages such as Thai, where publicly available telephony corpora are extremely limited and multilingual pretrained models allocate only a small fraction of their capacity to each language~\citep{babu2021xlsr,pratap2023scalingspeechtechnology1000,phatthiyaphaibun2022thaiwav2vec,aung-etal-2024-thonburian}. Addressing these gaps is essential for enabling accurate, reliable, and scalable AI-driven voice applications in enterprise contact centers.

In this work, we present a practical approach for rapidly adapting large-scale multilingual ASR models to telephony applications. Because no suitable public telephony corpus exists for Thai, we additionally designed a lightweight in-house data collection pipeline that records speech from recruited participants across diverse handsets, locations, and background conditions, yielding 77~hours of general telephony audio and an additional 104~hours focused on business-critical names and addresses (details in Sections~\ref{sec:telephony} and~\ref{sec:jargon}). Using this data, we fine-tune the open-source NVIDIA Canary 180M Flash and Canary 1B Flash models~\citep{canary_180m_flash_model_card,canary_1b_flash_model_card,puvvada2024canary} with the NVIDIA NeMo framework~\citep{nemo2019,nemo_github}.

To systematically evaluate the effectiveness of our approach, we design four targeted experiments: (1) language adaptation for Thai, (2) robustness to telephony-specific noise and distortions, (3) adaptation to domain-specific terminology such as names and addresses, and (4) latency optimization for real-time deployment. Performance is measured using character error rate (CER) for transcription accuracy and real-time factor (RTFx) for inference efficiency.

Our main findings can be summarized as follows: (i) fine-tuning on in-domain telephony data reduces CER on BOTNOI telephony evaluation from 23.31\% to 9.04\%, a 61\% relative improvement over the language-only baseline, and improvements of 54\% and 33\% over the commercial ElevenLabs Scribe v2 and Google Chirp 3 systems, respectively; (ii) a dedicated names-and-addresses adaptation stage further lowers CER on business-critical utterances from 16.98\% to 3.78\%; (iii) Canary 180M Flash closely matches the accuracy of the approximately 6× larger Canary 1B Flash while running at RTFx > 600, making it the preferred choice for production; and (iv) these gains are obtained with a single NVIDIA A100 GPU, demonstrating the feasibility of adapting large multilingual ASR models to specialized domains and resource-constrained languages with minimal computational overhead.

\section{Model and Framework Overview}

Modern end-to-end ASR is commonly organized around three broad modeling paradigms: Connectionist Temporal Classification (CTC)~\citep{graves2006connectionist}, transducer models~\citep{graves2012rnnt}, and attention-based encoder--decoder models~\citep{chan2016las}---all of which learn acoustic and linguistic representations jointly from paired speech--text data. On the encoder side, Transformer and Conformer architectures have become dominant due to their scalability and robustness across diverse datasets and languages~\citep{Vaswani+2017,conformer2020}. Self-supervised pretraining methods such as wav2vec~2.0~\citep{baevski2020wav2vec2}, HuBERT~\citep{hsu2021hubert}, and XLS-R~\citep{babu2021xlsr} further reduce the labeled-data requirement for downstream fine-tuning, while large-scale weakly supervised systems such as Whisper~\citep{whisper2022} and MMS~\citep{pratap2023scalingspeechtechnology1000} have shown that training on massive multilingual and noisy corpora yields strong robustness to accents, domain shift, and background noise.

The NVIDIA Canary family of models builds on these advances and represents a recent step toward unified multilingual ASR and speech-to-text translation~\citep{puvvada2024canary,canary_180m_flash_model_card,canary_1b_flash_model_card}. Canary adopts an encoder--decoder architecture that pairs a FastConformer encoder~\citep{rekesh2023fastconformer} with a Transformer-based decoder. FastConformer is a redesigned variant of the Conformer~\citep{conformer2020} that uses 8$\times$ subsampling and linearly scalable attention to achieve roughly 2.8$\times$ faster inference while retaining, and often improving, the accuracy of the original Conformer---properties that are particularly valuable for real-time telephony workloads~\citep{rekesh2023fastconformer,nemo_asr_models_docs}. The hybrid design leverages convolutional modules for local acoustic pattern modeling and self-attention for long-range dependency modeling, yielding strong performance across ASR, speech translation, and spoken language understanding.

Notably, Canary demonstrates that carefully curated, moderate-scale training data combined with multitask objectives can match or exceed systems trained on web-scale corpora, while keeping the parameter budget small~\citep{puvvada2024canary}. The Flash variants---Canary 180M Flash and Canary 1B Flash---are further optimized for low-latency inference, offering a favorable trade-off between model size, accuracy, and speed that is critical for voicebots and contact center automation~\citep{canary_180m_flash_model_card,canary_1b_flash_model_card}.

To support efficient development and deployment, we build on the NVIDIA NeMo framework~\citep{nemo2019,nemo_github,nemo_asr_models_docs}, an open-source toolkit for conversational AI that provides modular components for ASR, built-in data augmentation such as SpecAugment~\citep{park2019specaugment}, and training optimizations including mixed precision and distributed multi-GPU execution. These capabilities enable rapid fine-tuning of large pretrained models on domain-specific datasets---such as Thai telephony speech---while maintaining computational efficiency on commodity hardware. In this work, we adopt Canary 180M Flash and Canary 1B Flash as our base models and leverage NeMo to perform targeted fine-tuning on telephony-grade audio and domain-specific data, enabling robust adaptation of large multilingual ASR models to real-world enterprise environments.

\section{Telephony Data Collection and Model Training Configuration}

\begin{figure}[!t]
  \centering
  \includegraphics[width=0.98\linewidth]{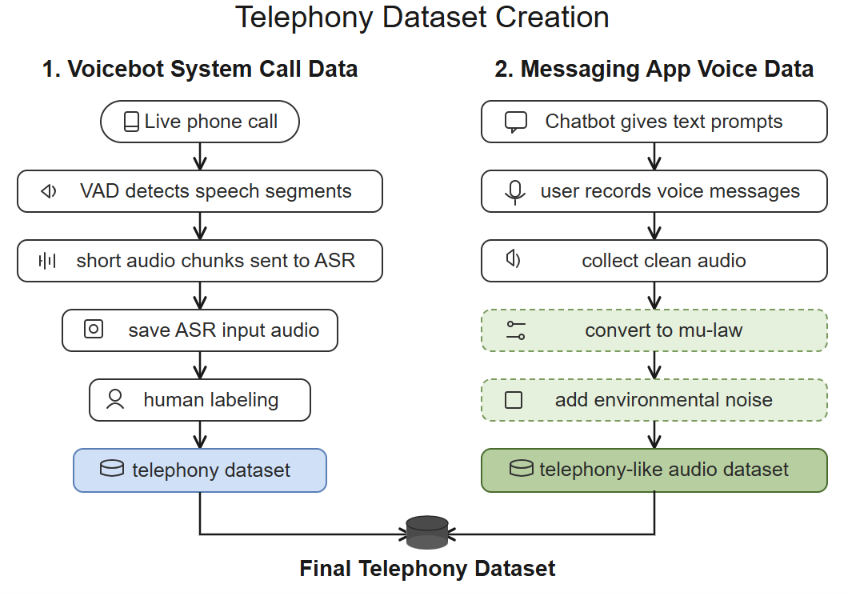}
  \caption{Overview of the telephony data collection pipeline, combining voicebot system call audio, prompted messaging-app recordings, and telephony-oriented augmentation.}
  \label{fig:telephony_data_collection}
\end{figure}

\subsection{Telephony data collection}
\label{sec:datacollection}
To construct a telephony-oriented speech dataset that reflects real deployment conditions while still scaling to the volume required for model adaptation, we combined live voicebot system audio with prompted speech collected through a messaging application. The overall collection pipeline is shown in \Cref{fig:telephony_data_collection}. This hybrid design lets us preserve the realism of genuine telephony channel conditions while also increasing data volume and lexical coverage through a more controllable collection process.

\subsubsection{Telephony data collection from the voicebot system}
The first branch draws directly on audio from our production voicebot telephone system, captured through interactions between consenting internal participants and the deployed agent over live phone calls. In this pipeline, user speech is automatically segmented into short utterances before transcription: the system applies Voice Activity Detection (VAD) to identify speech regions and forwards each detected segment to the ASR module. Because these speech chunks are already generated during normal voicebot operation, we directly save the same audio segments that are sent to ASR.

Each saved segment is then manually annotated with its corresponding transcription and incorporated into the dataset. This collection method provides highly authentic telephony audio, as it naturally captures the acoustic and channel conditions present in genuine telephone interactions. These include telephony channel characteristics, spontaneous speaking behavior, and realistic conversational pacing. Accordingly, this branch forms the core source of real-world telephony speech in our dataset, contributing a total of 25 hours of authentic recorded speech.

\subsubsection{Prompted speech collection via messaging app}
The second branch trades some acoustic realism for scale and lexical control. Here, a chatbot sends predefined prompts instructing users what to say, and users respond by recording voice messages through a messaging application. Because the spoken content is determined by the prompt, the transcript is known in advance and no manual labeling is required for these recordings.

This approach enables efficient collection of clean speech data with strong control over lexical content and domain coverage. Compared with real call collection, this method is substantially more scalable and is therefore well suited for gathering large volumes of speech within a limited time. However, because the resulting audio is recorded in a cleaner environment and through a different channel, it does not fully match the acoustic characteristics of real telephony speech. For this reason, an additional augmentation process is applied before the data is used for telephony ASR training.

\subsubsection{Telephony-oriented audio augmentation}
To make the prompted speech recordings more similar to real telephone audio, we applied a two-step augmentation process. First, the clean audio collected from the messaging application was converted to mu-law format, which simulates the compression characteristics commonly used in telephony systems. This step reduces the mismatch between clean voice-message recordings and actual telephone-channel audio.

Second, environmental noise was added to the mu-law audio to simulate realistic deployment conditions. The noise sources included both publicly available recordings and internally collected noise samples. Through this augmentation process, the prompted recordings were transformed into telephony-like speech that more closely resembles the acoustic conditions encountered in production voicebot calls.

Overall, the final telephony dataset is constructed from two complementary sources: authentic real-call data collected through the voicebot system and scalable prompted speech collected through a messaging application and converted into telephony-like audio through augmentation. This design allows us to balance realism, controllability, and collection efficiency in the dataset construction process, yielding 52 hours of noise-augmented training data.

All data was collected internally with consent; no external customer data was used. The real-call audio was recorded by internal participants interacting with the production voicebot over the live telephone system. Prompted recordings and the names-and-addresses set were produced by internal personnel and hired data-collection staff who were informed of the intended use of their recordings. The names and addresses themselves were synthetically generated, not drawn from real individuals.

\subsection{Model training configuration}
Following the Canary training tutorial notebook provided in the NeMo repository, the model was fine-tuned using the following specifications:
\begin{itemize}
  \item \textbf{Training framework:} NeMo \texttt{EncDecMultiTaskModel()} architecture
  \item \textbf{Hardware configuration:} Single NVIDIA A100 GPU (40GB VRAM)
  \item \textbf{Optimization settings:} AdamW optimizer; inverse square root annealing learning-rate schedule; mixed precision (fp16)
  \item \textbf{Dataset size / schedule:} 1 million training steps with an audio bucket size of 320 seconds
\end{itemize}

\section{Experiments and Results}
We conducted four experiments to evaluate the fine-tuning capability of the Canary models in handling multiple languages, telephony data, business-specific jargon, and latency. Across these experiments, we compare our fine-tuned Canary models against several baselines, including Whisper-large-v3~\citep{whisper2022,whisper_large_v3_model_card} as well as the commercial ElevenLabs Scribe v2 and Google Chirp 3 systems. In Experiments ~\ref{sec:language}--\ref{sec:jargon}, we fine-tuned Canary 180M Flash. In Experiment ~\ref{sec:latency}, we benchmarked Canary 180M Flash against Canary 1B Flash to examine the trade-off between accuracy and latency. We report character error rate (CER) as the primary metric and RTFx (real-time factor) for inference latency. We use CER rather than WER because Thai script does not delimit words with spaces, so word-level segmentation requires an additional, non-standardized tokenization step; CER avoids this dependency and is therefore the more consistent metric for Thai ASR evaluation.

\subsection{Datasets}
\label{sec:datasets}
We use three datasets across our experiments, summarized in Table~\ref{tab:datasets}.

\textbf{Thai public corpora.} To adapt the multilingual base model to Thai, we assembled 1{,}593 hours of Thai voice data from multiple publicly available sources, used for training throughout Sections~\ref{sec:language}--\ref{sec:latency}. Language-adaptation evaluation (Section~\ref{sec:language}) uses Common Voice 23 and Fleurs.

\textbf{BOTNOI telephony dataset.} Since no publicly available Thai telephony dataset exists, we constructed a dedicated dataset combining 25 hours of annotated real-call audio from our production voicebot system with 52 hours of prompted speech collected via messaging app and augmented to simulate telephony channel conditions (Section~\ref{sec:datacollection}), yielding 77 hours in total. We allocated 90\% for training and reserved 10\% for evaluation (Section~\ref{sec:telephony}).

\textbf{BOTNOI names and addresses dataset.} To evaluate adaptation to business-critical jargon, we generated a diverse list of Thai names using large language models and compiled addresses including building names, street names, and provinces. We recruited speakers to pronounce these over telephone connections from various locations, capturing a broad range of environmental noise conditions, yielding 104 hours of data. We allocated 90\% for training and reserved 10\% for evaluation (Section~\ref{sec:jargon}).

\begin{table}[h]
\centering
\caption{Summary of datasets used for fine-tuning and evaluation.}
\label{tab:datasets}
\begin{tabular}{llrl}
\toprule
Dataset & Source & Hours \\
\midrule
Thai public corpora & Multiple public sources & 1{,}593 \\
BOTNOI telephony (real-call) & Voicebot system calls, annotated & 25 \\
BOTNOI telephony (augmented) & Prompted speech + augmentation & 52 \\
BOTNOI names \& addresses & Prompted telephony recordings & 104 \\
\bottomrule
\end{tabular}
\end{table}

\subsection{Fine-tuning to adapt to local languages}
\label{sec:language}
We first evaluate whether the Canary fine-tuning recipe can effectively transfer to a local language setting. Because enterprise deployments in Southeast Asia often require strong support for underrepresented languages, this experiment focuses on adapting the multilingual base model to Thai and verifying that the same approach also generalizes to Vietnamese. We fine-tune on the Thai public corpora and evaluate on Common Voice 23 and Fleurs, as described in Section~\ref{sec:datasets}.

\begin{table}[H]
  \centering
  \caption{Result of fine-tuned Canary 180M Flash model with publicly available Thai datasets compared to open-source and commercial baselines.}
  \label{tab:thai_results}
  \begin{tabular}{lcc}
    \toprule
    Model & Common Voice 23 CER (\%) $\downarrow$ & Fleurs CER (\%) $\downarrow$ \\
    \midrule
    Whisper-large-v3 & 7.25 & 8.85 \\
    ElevenLabs Scribe v2 & 1.92 & 6.23 \\
    Google Chirp 3 & 3.46 & 10.29 \\
    BOTNOI Canary & 4.66 & 9.77 \\
    \bottomrule
  \end{tabular}
\end{table}

\paragraph{Summary}
Our fine-tuned Canary 180M Flash achieves competitive Thai CER, outperforming Whisper-large-v3 on both Common Voice 23 and Fleurs. While ElevenLabs Scribe v2 achieves lower CER on this clean read-speech evaluation (and Google Chirp 3 on Common Voice 23), our model's advantage emerges under telephony conditions (Section ~\ref{sec:telephony}), which are the target deployment setting. This confirms the feasibility of adapting large-scale multilingual models to local languages with strong accuracy, making them highly applicable for real-world business use cases such as transcription services and agent-assist systems. 

We also verified that the same recipe transfers to Vietnamese---training on roughly 1{,}800 hours of publicly available Vietnamese data yields CER comparable to open-source baselines on Common Voice 23---while the telephony and domain-specific experiments in the remainder of this report focus on Thai, where the BOTNOI telephony data is available.

\subsection{Fine-tuning to adapt to telephony data}
\label{sec:telephony}
We fine-tuned on the BOTNOI telephony dataset (Section~\ref{sec:datasets}), allocating 90\% for training and reserving 10\% for evaluation, and included the 1{,}593 hours of Thai language data from Section~\ref{sec:datasets} in training to retain general Thai language capability.

\begin{table}[H]
  \centering
  \caption{Result of fine-tuned Canary 180M Flash with BOTNOI telephony data compared to baseline ASR systems.}
  \label{tab:telephony_results}
  \begin{tabular}{lc}
    \toprule
    Model &  BOTNOI telephony \\
          &  CER (\%) $\downarrow$ \\
    \midrule
    Whisper-large-v3 & 48.44 \\
    ElevenLabs Scribe v2 & 19.68 \\
    Google Chirp 3 & 13.42 \\
    BOTNOI Canary (non-telephony) & 23.31 \\
    BOTNOI Canary (telephony) & 9.04 \\
    \bottomrule
  \end{tabular}
\end{table}

\paragraph{Summary}
Fine-tuning with telephony data substantially improved performance on telephony audio.

\subsection{Fine-tuning to adapt to domain-specific jargon}
\label{sec:jargon}
Although the results improved for telephony data, we found that the models still faced challenges in accurately transcribing business domain jargon. To address this, we conducted an additional experiment focused on names and addresses, aiming to evaluate whether domain-specific fine-tuning could enhance accuracy in these areas.

We fine-tuned on the BOTNOI names and addresses dataset (Section~\ref{sec:datasets}), allocating 90\% for training and reserving 10\% for evaluation, and incorporated the additional 1{,}593 hours of Thai language data from Section~\ref{sec:datasets} to preserve general Thai language capability.

\begin{table}[H]
  \centering
  \caption{Result of fine-tuned Canary 180M Flash model with names and addresses data compared to open-source and commercial baselines.}
  \label{tab:jargon_results}
  \small
  \begin{tabular}{lcc}
    \toprule
    Model & Names \& Addresses & BOTNOI telephony \\
          & CER (\%) $\downarrow$ & CER (\%) $\downarrow$ \\
    \midrule
    Whisper-large-v3 & 22.99 & 48.44 \\
    ElevenLabs Scribe v2 & 14.76 & 19.68 \\
    Google Chirp 3 & 9.86 & 13.42 \\
    BOTNOI Canary (telephony) & 16.98 & 9.04 \\
    BOTNOI Canary (telephony + names/addresses) & 3.78 & 10.89 \\
    \bottomrule
  \end{tabular}
\end{table}

\paragraph{Summary}
Domain-focused fine-tuning improved accuracy on names and addresses, demonstrating the ability to adapt to business-specific jargon. The slight increase in general telephony CER (from 9.04\% to 10.89\%) is an expected trade-off, reflecting a shift in the model's specialization toward names and addresses domain data.

\subsection{Latency}
\label{sec:latency}
Following the Open ASR Leaderboard convention~\citep{srivastav2025openasrleaderboardreproducible}, we report RTFx as the ratio of audio duration to processing time, so higher is faster; this is the inverse of the classical RTF, where lower is better.

Since NVIDIA released two Canary Flash models—180M and 1B—we aimed to evaluate the performance trade-offs between accuracy and latency. Our expectation was that the larger model would deliver higher accuracy, though potentially at the cost of increased latency.

Using the dataset from Experiment ~\ref{sec:jargon}, we fine-tuned the Canary 1B Flash model and reported both CER (Character Error Rate) and RTFx (Real-Time Factor) to compare its performance against the Canary 180M Flash. As reference points for inference speed, we include two Whisper-family models. The first, denoted BOTNOI Distil-whisper-large-v3, is a distilled Whisper model that we fine-tuned on the same telephony data as an internal reference point. The second is the pretrained Whisper-large-v3 (Radford et al., 2022; OpenAI, 2026), evaluated zero-shot. Because pretrained Whisper-large-v3 is not adapted to Thai telephony, its CER is not directly comparable to the fine-tuned models and is included only to illustrate the accuracy–speed envelope of the Whisper architecture on this task. Testing used a single 40GB A100 GPU with a batch size of 32. Note that these CER values are reported on Common Voice 23 for the final telephony-adapted checkpoints, and therefore differ from the language-adaptation-only results in Table ~\ref{tab:thai_results}.

\begin{table}[H]
  \centering
  \caption{CER and RTFx comparison of the Canary Flash models and baselines on Common Voice 23 Thai.}
  \label{tab:latency_results}
  \begin{tabular}{lcc}
    \toprule
    Model & CER (\%) $\downarrow$ & RTFx $\uparrow$ \\
    \midrule
    Whisper-large-v3 & 7.25 & 56.5 \\
    BOTNOI Distil-whisper-large-v3 & 3.94 & 75.4 \\
    BOTNOI Canary 180M Flash & 2.87 & 601.8 \\
    BOTNOI Canary 1B Flash & 2.73 & 531.4 \\
    \bottomrule
  \end{tabular}
\end{table}

\paragraph{Summary}
Canary 1B Flash achieved slightly higher accuracy but with slower inference speed. Given limited statistical significance and deployment constraints, we selected Canary 180M Flash for production due to its lower memory footprint and efficiency in real-time applications.

Our approach demonstrates how rapid fine-tuning with NVIDIA NeMo and open speech models translates seamlessly to production deployments where enterprises are able to scale voicebot and agent-assist solutions with robust, low-latency inference.

\section{Conclusion}
This report shows that NVIDIA Canary Flash models can be adapted effectively to enterprise telephony ASR through targeted fine-tuning with NVIDIA NeMo. A key part of this approach is our in-house telephony data collection pipeline, which provides fine-tuning data from real call audio and targeted business-domain speech. Across our experiments, the fine-tuned models remain competitive with Whisper-large-v3 on local-language evaluation and deliver substantial gains on in-domain telephony and business-specific speech. In particular, telephony adaptation reduces CER from 23.31\% to 9.04\%, while additional fine-tuning on names and addresses further reduces CER from 16.98\% to 3.78\%, demonstrating the importance of domain-specific data for production voicebot settings.

Our latency results further indicate that both Canary Flash variants are practical for real-time deployment, with Canary 1B Flash achieving the best accuracy and Canary 180M Flash offering a more favorable efficiency--accuracy trade-off for production use. Taken together, these findings suggest that moderate-scale, carefully collected telephony data---combined with multilingual pretrained models and efficient fine-tuning---can provide a strong foundation for robust ASR in contact center applications. 

\section{Limitations}
Our evaluation has several limitations. First, the telephony and domain-specific experiments are evaluated entirely on internally collected Thai data; no external Thai telephony benchmark exists for cross-validation, so absolute CER values may not transfer directly to other deployments. Second, the real-call audio was recorded by internal participants rather than from external callers, which faithfully reproduces telephony channel conditions but may not fully capture the speaking styles and spontaneity of genuine customer interactions. Third, the names-and-addresses adaptation improves domain accuracy at the cost of a small regression on general telephony speech (9.04\% to 10.89\% CER), indicating that a single model specialized for one domain trades off against broader robustness. Finally, our latency comparison reflects a single hardware and batch-size configuration; real-time factor may vary under different serving conditions.


\bibliography{colm2024_conference}
\bibliographystyle{colm2024_conference}


\end{document}